\documentstyle[prl,aps,twocolumn]{revtex}

\begin{document}

\twocolumn[\hsize\textwidth\columnwidth\hsize\csname
@twocolumnfalse\endcsname
\title{The Exact Solution of 1-D $SU(n)$ Hubbard Model}
\author{Boyu Hou$^1$ \thanks{E-mail:byhou@phy.nwu.edu.cn}\hspace{5mm}
      Dantao Peng$^1$ \thanks{E-mail:dtpeng@phy.nwu.edu.cn}\hspace{5mm}
      Ruihong Yue$^{\;a 1,2}$ \thanks{E-mail:yue@phy.nwu.edu.cn}\\[.3cm]
      $^a$CCAST (World Laboratory), P.O.Box 8730,\\ 
       Beijing, 100080, P.R.China\\
      $^1$Institute of Modern Physics Northwest University\\
       Xi'An,710069, P.R.China\\
      $^2$Institute of Theoretical Physics, Academica Science\\
       Beijing, 100080, P.R.China}
\date{}
\maketitle
\begin{abstract}
In this paper we derive out the exact solution of the $SU(n)$ Hubbard
model through the coordinate and the algebraic Bethe ansatz methods. The
energy spectrum and the Bethe ansatz equations are obtained. Furthemore, 
we analyse the ground state and give out the exact analytic solution of 
the model.


\end{abstract}]

\bigskip

\setcounter{equation}{0}

\section{Introduction}

\indent

The 1-D Hubbard model is one of the significant models in the study of the 
strongly correlated electron systems which might reveal an enlightening
role in understanding the mysteries of the high-$T_c$ superconductivity.
It also favours a lot of properties of integrable models in
non-pertubative quantum field theory and mathematical physics. It's exact
solution was first given by Lieb and Wu\cite{Lieb}. Based on the Bethe
ansatz equations (Lieb-Wu's equations), the 1-D Hubbard model was
extensively discussed in
Refs.\cite{AO,GV,CF,FW,AAJ,MT1,HS,FK,MT2,TK,NTA,HV,JPA}. Although there
are lots of works on Hubbard model, the integrability was finished until
1986 by Shastry\cite{Shastry}, Olmedilla and Wadati \cite{OW} in both
boson and fermion grade versions. Moreover, the eigenvalue of the transfer
matrix related to the Hubbard model was suggested in Ref.\cite{Shastry}
and proved through different method in Ref.\cite{RT,Martins2}. Besides,
the integrability and the exact solution of the  1-D Hubbard model with
open boundary condition have been investigated by several 
authors.\cite{HM,TR,SW}.

Recently, based on the Lie algebra knowledge, Maassarani and Mathieu
constructed the Hamiltonian of the $SU(n)$ XX model and showed its 
integrability\cite{MM}. Considering two coupled $SU(n)$ XX models,
Maassarani succeeded in generalizing Shastry's method to construct the
$SU(n)$ Hubbard model\cite{Ma1}. Furthermore, he found the related 
$R$-matrix which ensures the integrability of the one-dimensional $SU(n)$
Hubbard model\cite{Ma2}.(It was also proved by Martins for $n = 3, 
4$.\cite{Martins1}, and by Yue and Sasaki\cite{yuesasaki} for general $n$
in terms of Lax-pair formalism.) The exact solution of the $SU(3)$ Hubbard
model was also given in Ref. \cite{HPY}. However, the eigenvalue and the
eigenvectors of the $SU(n)$ Hubbard model have not been given yet.

In this paper, we apply the same method as done in Ref.\cite{HPY} to find
the exact solution of the 1-D $SU(n)$ Hubbard model. After recalling some
basic notations, we write down the wave functions and derive out the
$2$-particle scattering matrix which governs the amplitude of the wave
functions in section 2. In section 3 with the help of the Yang-Baxter
relation we employ the algebraic Bethe ansatz method to discuss the
amplitude of wave functions and the Bethe ansatz equations. The exact
solution of the $SU(n)$ Hubbard model was then given out. Under the
thermodynamical limit, the explicitly analytic form of the ground state
energy is given in section 4. In section 5, we make some conclusions and
list some questions to be considered.

\section{The Coordinate Bethe Ansatz}

\indent

The Hamiltonian of the $SU(n)$ Hubbard model is
\begin{eqnarray}
\label{Hamiltonian}
H & = & \sum_{k=1}^{L}\sum_{\alpha=1}^{n-1}(E_{\sigma,  k}^{n \alpha}
    E_{\sigma, k+1}^{\alpha n} + E_{\sigma, k}^{\alpha n}
    E_{\sigma, k+1}^{n \alpha} + (\sigma \rightarrow \tau))\nonumber\\
  & & + \frac{U n^2}{4}\sum_{k=1}^{L}C_{\sigma, k}C_{\tau, k},
\end{eqnarray}
where $U$ is the coulumb coupling constant, and $ E_{a, k}^{\alpha \beta}
(a = \sigma, \tau)$ is a matrix with zeros everywhere except for an one at
the intersection of row $\alpha$ and column $\beta$:
\begin{equation}
(E^{\alpha \beta})_{l m} = \delta_{l}^{\alpha}\delta_{m}^{\beta},
\end{equation}
The subscripts $a$ and $k$ stand for two different $E$ operators at site 
$k(k =1,\cdots,L)$. The $n \times n$ diagonal matrix $C$ is defined by $C
= \sum_{\alpha < n}E^{\alpha \alpha} - E^{n n}$. We have also assumed the
periodic boundary condition $E^{\alpha \beta}_{k+L} = E^{\alpha \beta}_k$.

It was shown that the Hamiltonian (\ref{Hamiltonian})  has a
$(su(n-1)\oplus u(1))_{\sigma}\oplus(su(n-1)\oplus u(1))_{\tau}$ symmetry.
The generators are
\begin{equation}
\label{generator}
J_{a}^{\alpha \beta} = \sum_{k=1}^{n}E_{a, k}^{\alpha \beta}
\end{equation}
and
\begin{equation}
K_a = \sum_{k=1}^L C_{a, k},
\hspace{.3cm} \alpha, \beta = 1,\cdots,n-1,
\hspace{.3cm} a = \sigma,\tau.
\end{equation}
It is worthy to point out that the system enjoys the $SO(4)$ symmetry when
$n=2$\cite{YS}. But the generator are different from Eq. (\ref{generator}).

Before proceeding the coordinate Bethe ansatz approach, we first begin by
introduce some notations as in Ref.\cite{HPY}. In $SU(n)$ Hubbard model,
there are two types of particles named $\sigma$ and $\tau$, and each
particle can occupy $(n-1)$ possible states. The same type of particles
cannot appear in one site, but two different types of particles can occupy
one same site. We denote $|n\rangle_j$ the vacuum state of $j$-th site,
$|1\rangle_j, \dots, |n-1\rangle_j$ the particle states of $j$-th site
respectively. Under the appropriate basis
\begin{equation}
|1\rangle_j = \left (
\begin{array}{c}
1\\
0\\
\vdots\\
0\\
\end{array} \right )_j,
|2\rangle_j = \left (
\begin{array}{c}
0\\
1\\
\vdots\\
0\\
\end{array} \right )_j,\cdots,
|n\rangle_j = \left (
\begin{array}{c}
0\\
0\\
\vdots\\
1\\
\end{array} \right )_j ,
\end{equation}
we can prove that $E^{\alpha n}_j$ and $E^{n \alpha}_j$ act as a creating
and a annihilating operators of $|\alpha\rangle_j$ state particle
respectively. 

Define the particle number operators of the particles labeled with ($s$,
$\alpha$) as:
\begin{eqnarray*}
\hat{N}_s^{\alpha} & = & \sum_{k=1}^L E_{s, k}^{\alpha n}E_{s, k}^{n
\alpha} = \sum_{k=1}^L E_{s, k}^{\alpha \alpha}, \nonumber\\
& & ~~~~(\alpha = 1, 2, \cdots, n-1, s = \sigma, \tau)
\end{eqnarray*}
we can easily prove that these $2(n-1)$ particle number operators 
commutate with the Hamiltonian (\ref{Hamiltonian}):
\begin{equation}
[\hat{N}_s^{\alpha}, H] = 0,
\end{equation}
and commute with each other. So such particle numbers are conserved.

Our main purpose is to give out the eigenvalues and the eigenstates of the
Hamiltonian (\ref{Hamiltonian}). In this section, we first give out the
two particle sacttering matrix of the system.
 
In the coordinate Bethe ansatz method, the eigenstates of the Hamiltonian
can be assumed as
\begin{eqnarray}
\label{wavefunction}
|\psi_{N_0}\rangle & = & \sum_{x_1\leq x_2\leq\cdots\leq x_{N_0} = 1}^L
f_{\sigma_1 \sigma_2 \cdots \sigma_{N_0}}^{\alpha_1 \alpha_2 \cdots
\alpha_{N_0}}(x_1, x_2, \cdots, x_{N_0})\nonumber\\
& & \times E_{\sigma_1 x_1}^{\alpha_1 n}E_{\sigma_2 x_2}^{\alpha_2
n}\cdots E_{\sigma_{N_0} x_{N_0}}^{\alpha_{N_0} n}|0\rangle,
\end{eqnarray}
here $|0\rangle$ is the vacuum state of the total chain and $|0\rangle =
\prod_{j=1}^L\otimes|n\rangle_j$ and
\begin{eqnarray}
f & = & \sum_{P, Q}\epsilon_P\epsilon_Q A_{\sigma_{Q_1}
\sigma_{Q_2} \cdots \sigma_{Q_{N_0}}}^{\alpha_{Q_1} \alpha_{Q_2} \cdots
\alpha_{Q_{N_0}}}(k_{P_{Q_1}}, k_{P_{Q_2}}, \cdots,
k_{P_{Q_{N_0}}})\nonumber\\
& & \times \theta(x_{Q_1}\leq x_{Q_2}\leq\cdots\leq x_{Q_{N_0}})\exp(i
\sum_{j = 1}^{N_0}k_{P_j}x_{Q_j}),
\end{eqnarray}
with
\begin{equation}
\theta(x_1\leq x_2\leq\cdots\leq x_{N_0}) = \left\{
\begin{array}{l}
1, \quad x_1 < x_2 < \cdots < x_{N_0},\\
\frac{1}{2}, \quad x_i = x_j ,\\
0, \quad others.
\end{array}\right .
\end{equation}
Here $\alpha_i = 1, 2,\cdots, n-1 (i=1,2,\cdots, N_0)$ stands for the
different particle states, $x_i$ the position of the particle, $\sigma_i =
\sigma, \tau$ the type of $i$th particle. The summation $P$ and $Q$ are
taken over all permutations of $N_0$ momenta $k_j$ and $N_0$ coordinates
$x_j$ respectively.The symbols $\epsilon_P$ and $\epsilon_Q$ are the
parities of two kinds of permutations. Substituting the wave function into
the Schr\"odinger equation
\begin{equation}
\label{Schrodinger}
H |\psi_{N_0}\rangle = E |\psi_{N_0}\rangle,
\end{equation}
we can get
\begin{eqnarray}
&&A_{\cdots, \sigma_i, \sigma_j, \cdots}^{\cdots, \alpha_i, \alpha_j,
\cdots}(\cdots, k_i, k_j, \cdots) = \nonumber\\
&&~~~S_{\sigma_i \sigma_j}^{\alpha_i
\alpha_j}(\sin k_i, \sin k_j)A_{\cdots, \sigma_j, \sigma_i, 
\cdots}^{\cdots, \alpha_j, \alpha_i, \cdots}(\cdots, k_j, k_i, \cdots).
\end{eqnarray}
with $S_{\sigma_i\sigma_i}^{\alpha_i\alpha_j}(\sin k_i, \sin k_j)$ being
the two-particle scattering matrix:
\begin{equation}
\label{scattering}
S_{\sigma_i \sigma_j}^{\alpha_i \alpha_j}(\sin k_i, \sin k_j) =
\frac{\sin k_i - \sin k_j + i \gamma P_{\sigma_i \sigma_j}^{\alpha_i
\alpha_j}}{\sin k_i - \sin k_j + i \gamma}
\end{equation}
where $\gamma = \frac{n^2 U}{2}$, $P_{\sigma_i \sigma_j}^{\alpha_i 
\alpha_j}$ is the direct product of two kinds of permutation operators
which permutes the particle styles and particle states simultaneously.
The energy of the Hamiltonian on this wave function is
\begin{equation}
E = 2 \sum_{i=1}^{N_0}\cos k_i + \frac{\gamma}{2}(L-2{N_0}).
\end{equation}

It is worthy to point out that the ansatz (\ref{wavefunction}) is not
complete \cite{FVK}, which can be observed from a simple example. The
state $|0\rangle$ and the states without any $|n\rangle_j$ are all the
eigenstates of the Hamiltonian (\ref{Hamiltonian}) with the same
eigenenergy $E_0 = \frac{\gamma}{2}L = \frac{n^2}{4}UL$, so for this
eigenvalue, the degeneracy of the eigenstate is $(n-1)^{2L} + 1$.

For the convenience, we denote by $\zeta_j$ the amplitude $A_{\cdots,
\sigma_j, \sigma_l, \cdots}^{\cdots, \alpha_j, \alpha_l, \cdots}(\cdots,
k_j, k_l, \cdots)$. When the $j$-th particle moves across the else 
particle, it gets an $S$-matrix $S_{jl}(q_j, q_l)$. Here $q_j = \sin
k_j$. The periodic boundary condition leads the following constrains on
the amplitude $\zeta_j$
\begin{equation}
\label{eigenvalue}
S_{j j+1}S_{j j+2}\cdots S_{j N_0}S_{j 1}S_{j 2}\cdots S_{j j-1}\zeta_j =
e^{i k_j L}\zeta_j.
\end{equation}
Eq.(\ref{eigenvalue}) is similar with the Yang's eigenvalue 
problem\cite{Yang}. Its solution will give out the Bethe ansatz equations
and the amplitude.

\section{The eigenvalue and the eigenstates}

\indent

In the above section, we have obtained the scattering matrix. It can be 
proved that it satisfies the Yang-Baxter equation\cite{Yang}
\begin{eqnarray}
&&S_{j l}(q_j, q_l)S_{j k}(q_j, q_k)S_{l k}(q_l, q_k) = \nonumber\\
&&~~~~~~~~~~S_{l k}(q_l, q_k)S_{j k}(q_j, q_k)S_{j l}(q_j, q_l).
\end{eqnarray}
Define the monodromy matrix as
\begin{equation}
\label{monodromy-matrix}
T_{\tau}(q)=S_{\tau 1}(q, q_1)S_{\tau 2}(q, q_2)\cdots S_{\tau N_0}(q,
q_{N_0}),
\end{equation}
here $\tau$ is the auxiliary space. We can easily prove that the monodromy
matrix satisfies the Yang-Baxter relation
\begin{equation}
S_{\tau \tau^\prime}(q, q^\prime)T_{\tau}(q)T_{\tau^\prime}(q^\prime) =
T_{\tau^\prime}(q^\prime)T_{\tau}(q)S_{\tau \tau^\prime}(q, q^\prime).
\end{equation}
So the eigenvalue problem (\ref{eigenvalue}) is equivalent to finding the
solution of following problem:
\begin{equation}
\label{tran-prob}
tr_{\tau}T_{\tau}(q_j)\zeta = X_j \zeta = \epsilon(q_j)\zeta.
\end{equation}
Here $\epsilon(q_j)$ and $\zeta$ are the eigenvalue and the eigenstate
respectively of the transfer matrix $X_j$ which is the matrix trace of the
monodromy matrix.

\subsection{The fundamental commutation rules}

\indent

From the point of view of a vertex model, we can interpret the matrix
$S_{\tau j}$ as the vertex operator, the matrix $S_{\tau \tau^\prime}$ as
the $R$-matrix. So the $R$-matrix is an $(2(n-1))^2\times (2(n-1))^2$
matrix
\begin{equation}
R_{j l}(q_j, q_l) = \alpha_2(q_j, q_l) + \alpha_3(q_j , q_l)P_{j l}
\end{equation}
where
\begin{equation}
\alpha_2(q_j , q_l) = \frac{q_j - q_l}{q_j - q_l + i\gamma},~
\alpha_3(q_j , q_l) = \frac{i\gamma}{q_j - q_l + i\gamma},
\end{equation}
and $P_{j l}$ is a $(2(n-1))^2\times (2(n-1))^2$ dimensional permutation
operator.

The vertex operator is
\begin{eqnarray}
{\cal L}_j(q - q_j) & = & S_{\tau j}(q , q_j) = \nonumber\\
& & \alpha_2 + \alpha_3 \sum_{\alpha, \beta = 1}^{2(n-1)}E^{\alpha
\beta}\otimes E^{\beta \alpha}_j,
\end{eqnarray}
where $E^{\alpha \beta}$ is a $2(n-1)\times 2(n-1)$ matrix with $0$
everywhere except for an $1$ at the intersection of raw $\alpha$ and
column $\beta$ and $E^{\alpha \beta}_j$ is an operator acting on the
$j$-th space.

The local reference state can be defined as:
$$
|0\rangle^{(1)}_j = \left (
\begin{array}{c}
1\\
0\\
\end{array}
\right )^{(1)}_{S_j} \otimes \left (
\begin{array}{c}
1\\
0\\
\vdots\\
0\\
\end{array}
\right )^{(1)}_{T_j~(n-1)\times 1} = \left (
\begin{array}{c}
1\\
0\\
\vdots\\
0\\
\end{array}
\right )^{(1)}_{j~2(n-1)\times 1},
$$
where $S_j$ and $T_j$ are the $j$-th particle style space and particle
state space respectively.

The action of $L$-operator on this state has the following property:
\begin{eqnarray}
\label{L-property}
{\cal L}_{\tau j}(q ) |0\rangle^{(1)}_j & = & \left (
\begin{array}{ccccc}
\alpha_1 & \alpha_3 E^{2 1}_j &\alpha_3 E^{3 1}_j & \cdots & \alpha_3
E^{2(n-1) 1}_j \\[3mm]
0 & \alpha_2 & 0 & \cdots & 0 \\[3mm]
0 & 0 & \alpha_2 & \cdots & 0 \\[3mm]
\vdots & \vdots & \vdots & \vdots & \vdots \\[3mm]
0 & 0 & 0 & \cdots & \alpha_2 \\[3mm] 
\end{array}\right )\nonumber\\
& & \times|0\rangle^{(1)}_j,
\end{eqnarray}
here $\alpha_1(q , q_j) = \alpha_2(q , q_j) + \alpha_3(q , q_j) = 1$.

The global reference state $|0\rangle^{(1)}$ is then defined by the tensor
product of local reference states, i.e. $|0\rangle^{(1)} =
\prod_{j=1}^L\otimes|0\rangle^{(1)}_j$. ( Here we use the superscript
$(1)$ to distinguish $|0\rangle^{(1)}$ with $|0\rangle$ appeared in the 
above section). The property of the L-operator suggests the representation
of the monodromy matrix to take the following form
\begin{equation}
T_\tau(q) = \left (
\begin{array}{ll}
B(q) & {\bf B}(q)\\
{\bf C}(q) & {\bf A}(q)\\
\end{array}\right )_{2(n-1)\times 2(n-1)}
\end{equation}
where ${\bf B}(q)$ and ${\bf C}(q)$ are vectors with dimensions $1 \times
(2n-3)$ and $(2n-3) \times 1$ respectively. The operator ${\bf A}(q)$ is a
$(2n-3) \times (2n-3)$ matrix and we denote its elements by $A_{a b}(q)$.
The remaining operator $B(q)$ is a scalar.

In the framework of the above partition the eigenvalue problem for the
transfer matrix (\ref{tran-prob}) becomes
\begin{equation}
\label{transferprob}
[B(q) + \sum_{a=1}^{2n-3} A_{a a}(q)]|\Phi\rangle = 
\Lambda(q)|\Phi\rangle.
\end{equation}
where $\Lambda(q)$ and $|\Phi\rangle$ correspond to the
eigenvalue and the eigenstate respectively. From Eq. (\ref{L-property}),
we know the action of the monodromy matrix on the reference state
$|0\rangle^{(1)}$:
\begin{eqnarray}
\label{monodromyaction}
&& B(q)|0\rangle^{(1)} = |0\rangle^{(1)},~~~~{\bf C}(q)|0\rangle^{(1)} =
0,\nonumber\\
&& A_{a b}(q)|0\rangle^{(1)} =\delta_{a b}\prod_{j=1}^{N_0}\alpha_2(q,
q_j)|0\rangle^{(1)},
\end{eqnarray}
where $a, b = 1, 2, \cdots, 2n-3$. The operator ${\bf B}(q)$ plays the
role of creation operators over the reference state $|0\rangle^{(1)}$.

To make further progress we have to recast the Yang-Baxter algebra in the
form of commutation relations for the elements of the monodromy matrix.
We shall start our discussion by the commutation rule between
the operators ${\bf B}(q)$ and ${\bf B}(p)$:
\begin{equation}
\label{comm-BvBv}
{\bf B}(q)\otimes{\bf B}(p) = [{\bf B}(p)\otimes{\bf
B}(q)]\cdot\hat{r}^{(1)}(q, p).
\end{equation}
where $\hat{r}^{(1)}(q, p)$ is an auxiliary $(2n-3)^2\times(2n-3)^2$
matrix given by
\begin{equation}
\label{rdef}
\hat{r}^{(1)}_{12}(q, p) = \alpha_3(q , p) + \alpha_2(q , p)P^{(1)}_{12}
\end{equation}
where $P^{(1)}_{12}$ is an $(2n-3)^2\times (2n-3)^2$ dimensional
permutation operator.

The commutation relations between the diagonal and creation operator ${\bf
B}(q)$ are
\begin{eqnarray}
\label{comm-ABv}
{\bf A}(q)\otimes{\bf B}(p) & = &  \frac{\alpha_1(q, p)}{\alpha_2(q,
p)}[{\bf B}(p)\otimes{\bf A}(q)]\cdot\hat{r}^{(1)}(q, p)\nonumber\\
& & - \frac{\alpha_3(q, p)}{\alpha_2(q, p)}{\bf B}(q)\otimes{\bf A}(p),
\end{eqnarray}
\begin{equation}
\label{comm-BBv}
B(q){\bf B}(p) = \frac{\alpha_1(p, q)}{\alpha_2(p, q)}{\bf B}(p)B(q) -
\frac{\alpha_3(p, q)}{\alpha_2(p, q)}{\bf B}(q)B(p).
\end{equation}

Now we have set up the basic tools to construct the eigenstates of the
eigenvalue problem (\ref{transferprob}). In the next section we will show
how this problem can be solved with the help of the commutations rules
(\ref{comm-BvBv}), (\ref{rdef}), (\ref{comm-ABv})-(\ref{comm-BBv}).

\subsection{The nested Bethe ansatz}

\indent

The purpose of this subsection is to solve the eigenvalue problem of the
transfer matrix (\ref{transferprob}).

First, we construct the eigenstates of the transfer matrix. The
eigenstates of the transfer matrix are in principle built up in terms of a
linear combination of the products of  many creation operators acting on
the reference state, which are characterized by a set of rapidities
parameterizing the creation operators. We define the arbitrary
$N_1$-particle eigenstate as:
\begin{equation}
\label{eigenvector}
|\Phi_{N_1}(\{p^{(1)}_l\})\rangle = {\bf \Phi}_{N_1}(p^{(1)}_1, p^{(1)}_2,
\cdots, p^{(1)}_{N_1})\cdot {\bf\cal F}|0\rangle^{(1)}
\end{equation}
where the mathematical structure of the eigenvector ${\bf 
\Phi}_{N_1}(p^{(1)}_1, p^{(1)}_2, \cdots, p^{(1)}_{N_1})$ will be
described in terms of the creation operators. We denote the components of
vector ${\bf \cal F}$ by ${\bf\cal F}^{a_1\cdots a_{N_1}}$ which will be
determined later on, where the index $a_i$ runs over $(2n-3)$ possible
values $a_i = 1, 2, \cdots, 2n-3$.

We construct the eigenvectors ${\bf \Phi}_{N_1}(p^{(1)}_1, \cdots,
p^{(1)}_{N_1})$ as
\begin{equation}
{\bf \Phi}_{N_1}(p^{(1)}_1, \cdots, p^{(1)}_{N_1}) = {\bf 
B}(p^{(1)}_1)\otimes{\bf \Phi}_{N_1 - 1}(p^{(1)}_2, \cdots,
p^{(1)}_{N_1}),
\end{equation}
here we have formally identified ${\bf \Phi}_0$ with the unity vector. 
This vector has the following symmetry:
\begin{eqnarray}
&&{\bf \Phi}_{N_1}(p^{(1)}_1, \cdots, p^{(1)}_{j-1}, p^{(1)}_j, \cdots,
p^{(1)}_{N_1}) = \nonumber\\
&&{\bf \Phi}_{N_1}(p^{(1)}_1, \cdots, p^{(1)}_j, p^{(1)}_{j-1},
\cdots, p^{(1)}_{N_1})\cdot\hat{r}^{(1)}_{j-1 j}(p^{(1)}_{j-1},
p^{(1)}_j),
\end{eqnarray}
where the subscripts in $\hat{r}^{(1)}_{j-1 j}(p^{(1)}_{j-1}, p^{(1)}_j)$
emphasize the positions in the $N_1$-particle space
$V_1\otimes\cdots\otimes V_{j-1}\otimes V_j\otimes\cdots\otimes V_{N_1}$
on which this matrix acts non-trivially. Here we have already assumed that
the $(N_1 - 1)$-particle state was already symmetrized. 

Applying the diagonal elements of monodromy matrix on this eigenstate, we
have
\begin{eqnarray}
\label{B-action}
&& B(q) |\Phi_{N_1}(\{p^{(1)}_l\})\rangle =
\prod_{i=1}^{N_1}\frac{\alpha_1(p^{(1)}_i, q)}{\alpha_2(p^{(1)}_i, q)} 
|\Phi_{N_1}(\{p^{(1)}_l\})\rangle\nonumber\\ 
&&~~~~~~~~~~~~~ - \sum_{i=1}^{N_1}\frac{\alpha_3(p^{(1)}_i,
q)}{\alpha_2(p^{(1)}_i,
q)}\prod_{k=1, k \neq i}^{N_1}\frac{\alpha_1(p^{(1)}_k,
p^{(1)}_i)}{\alpha_2(p^{(1)}_k, p^{(1)}_i)}\nonumber\\
&&~~~~~~~~~~~~~ \times|\Psi_{N_1 - 1}^{(1)}(q, p^{(1)}_i; \{p^{(1)}_l\}),
\end{eqnarray}
\begin{eqnarray}
\label{A-action}
&& \sum_{a=1}^{2n-3} A_{a a}(q)|\Phi_{N_1}(\{p^{(1)}_l\})\rangle = 
\prod_{j=1}^{N_0}\alpha_2(q, q_j)\nonumber\\
&&~~~~~ \times \prod_{i=1}^{N_1}\frac{\alpha_1(q, p^{(1)}_i)}{\alpha_2(q,
p^{(1)}_i)}\Lambda^{(1)}(q, \{p^{(1)}_l\}) 
|\Phi_{N_1}(\{p^{(1)}_l\})\rangle\nonumber\\
&&~~~~~ - \sum_{i=1}^{N_1}\frac{\alpha_3(q, p^{(1)}_i)}{\alpha_2(q,
p^{(1)}_i)}\prod_{j=1}^{N_0}\alpha_2(p^{(1)}_i, q_j)\prod_{k=1, k \neq 
i}^{N_1}\frac{\alpha_1(p^{(1)}_i, p^{(1)}_k)}{\alpha_2(p^{(1)}_i,
p^{(1)}_k)}\nonumber\\
&&~~~~~ \times \Lambda^{(1)}(p^{(1)}_i, \{p^{(1)}_l\})|\Psi_{N_1 -
1}^{(1)}(q, p^{(1)}_i, \{p^{(1)}_l\})\rangle,
\end{eqnarray} 
where
\begin{eqnarray}
\label{func-def}
&&|\Psi_{N_1 - 1}^{(1)}(q, p^{(1)}_i, \{p^{(1)}_l\})\rangle = \nonumber\\
&&~~~~~ {\bf B}(q)\otimes {\bf \Phi}_{N_1 - 1}(p^{(1)}_1, \cdots,
\check{p}^{(1)}_i, \cdots, p^{(1)}_{N_1})\nonumber\\
&&~~~~~ \times \hat{O}_i^{(1)}(p^{(1)}_i, \{p^{(1)}_l\})\cdot{\bf\cal
F}|0\rangle^{(1)}.
\end{eqnarray}
and
\begin{equation}
\label{Ofunc-def}
\hat{O}_i^{(1)}(p^{(1)}_i, 
\{p^{(1)}_k\})=\prod_{k=1}^{i-1}\hat{r}^{(1)}_{k, k+1}(p^{(1)}_k,
p^{(1)}_i).
\end{equation}
The symbol $\check{p}^{(1)}_i$ means that the rapidity $p^{(1)}_i$ is
absent from the set $\{p^{(1)}_1, \cdots, p^{(1)}_{N_1}\}$.

The terms proportional to the $N_1$-particle eigenstate 
$|\Phi_{N_1}(\{p^{(1)}_l\})\rangle$ are denominated the wanted terms
because they contribute directly to the eigenvalue. From Eqs.
(\ref{B-action})-(\ref{Ofunc-def}), we can directly get the eigenvalue of
the $N_1$-particle state
\begin{eqnarray}
\label{tran-value}
&& \Lambda(q, \{p^{(1)}_l\}) = \prod_{i=1}^{N_1}\frac{\alpha_1(p^{(1)}_i,
q)}{\alpha_2(p^{(1)}_i, q)}\nonumber\\
&&~~~~~~~~ + \prod_{j=1}^{N_0}\alpha_2(q, q_j) 
\prod_{i=1}^{N_1}\frac{\alpha_1(q, p^{(1)}_i)}{\alpha_2(q,
p^{(1)}_i)}\Lambda^{(1)}(q, \{p^{(1)}_l\}).
\end{eqnarray}
The remaining ones are called unwanted terms and can be eliminated by
imposing further restrictions which are known as the Bethe ansatz
equations:
\begin{eqnarray}
\label{tran-BA}
\prod_{j=1}^{N_0}\frac{\alpha_1(p^{(1)}_i, q_j)}{\alpha_2(p^{(1)}_i,
q_j)} & = & \prod_{k=1, k \neq i}^{N_1}\frac{\alpha_2(p^{(1)}_k, 
p^{(1)}_i)}{\alpha_2(p^{(1)}_i, p^{(1)}_k)}\Lambda^{(1)}(p^{(1)}_i, 
\{p^{(1)}_l\}),\nonumber\\
&&~~~~~~~~~~~~~~~i = 1, 2, \cdots, N_1.
\end{eqnarray}

In fact, the undeterminated eigenvalue $\Lambda^{(1)}(q, \{p^{(1)}_l\})$
must satisfy the following auxiliary problem
\begin{equation}
\label{auxpro}
T^{(1)}(q, \{p^{(1)}_l\})^{b_1\cdots b_{N_1}}_{a_1\cdots a_{N_1}}{\cal
F}^{b_1\cdots b_{N_1}} = \Lambda^{(1)}(q, \{p^{(1)}_l\}){\cal 
F}^{a_1\cdots a_{N_1}},
\end{equation}
where the inhomogeneous transfer matrix $T^{(1)}(q, \{p^{(1)}_l\})$ is
\begin{eqnarray}
&& T^{(1)}(q, \{p^{(1)}_l\})^{b_1\cdots b_{N_1}}_{a_1\cdots a_{N_1}} =
\hat{r}^{(1)}(q, p^{(1)}_1)^{c_1 b_1}_{a_1 d_1}\nonumber\\
&& ~~~~~~~~~~~ \times \hat{r}^{(1)}(q, p^{(1)}_2)^{d_1 b_2}_{a_2
d_2}\cdots\hat{r}^{(1)}(q,p^{(1)}_{N_1})^{d_{N_1 - 1} b_{N_1}}_{a_{N_1}
c_1}.
\end{eqnarray}
This auxiliary eigenvalue problem has the same structure as the eigenvalue
problem of the transfer matrix, so we can write the solution of this
auxiliary eigenvalue problem from the foregoing procedure directly.

The eigenvalue of this auxiliary eigenvalue problem is
\begin{eqnarray}
\label{Aux-value}
&& \Lambda^{(1)}(q, \{p^{(1)}_l\}, \{p_k^{(2)}\}) = \prod_{i=1}^{N_2} 
\frac{\alpha_1(p^{(2)}_i, q)}{\alpha_2(p^{(2)}_i, q)}\nonumber\\
&&~~~~~~ + \prod_{j=1}^{N_1}\alpha_2(q, p^{(1)}_j) 
\prod_{i=1}^{N_2}\frac{\alpha_1(q, p^{(2)}_i)}{\alpha_2(q,
p^{(2)}_i)}\Lambda^{(2)}(q, \{p^{(2)}_l\}).
\end{eqnarray}
and the parameters $\{p^{(2)}_i\}$ satisfy the following Bethe ansatz
equations:
\begin{eqnarray}
\label{Aux-BA}
\prod_{j=1}^{N_1}\frac{\alpha_1(p^{(2)}_i, p^{(1)}_j)}{\alpha_2(p^{(2)}_i,
p^{(1)}_j)} & = & \prod_{k=1, k \neq i}^{N_2}\frac{\alpha_2(p^{(2)}_k,
p^{(2)}_i)}{\alpha_2(p^{(2)}_i, p^{(2)}_k)}\nonumber\\
& & \times \Lambda^{(2)}(p^{(2)}_i, \{p^{(1)}_l\},
\{p^{(2)}_k\}),\nonumber\\
& & ~~~~~~~~ i = 1, \cdots, N_2. 
\end{eqnarray}
Here $\Lambda^{(2)}(q, \{p^{(1)}_l\}, \{p^{(2)}_k\})$ is determined by the
second auxiliary eigenvalue problem.

Comparing with above procedure, we know that this second auxiliary
eigenvalue problem is equivalent to finding the eigenvalue of the transfer
matrix of the $SU(n-1)$ Hubbard model. Applying the foregoing procedure
repeatedly until the dimension of the lower level space becomes equal to
one, at this point, the final auxiliary $r$-matrix is 
\begin{equation}
\hat{r}^{(2n-3)}(q, p) = 1,
\end{equation}
and the nested Bethe ansatz closed. For $n=2$ there is no nested 
necessary.

Now we have obtained the eigenvalue of the transfer matrix:
\begin{eqnarray}
&& \Lambda(q, \{p^{(1)}_l\}) = \prod_{i=1}^{N_1}\frac{\alpha_1(p^{(1)}_i,
q)}{\alpha_2(p^{(1)}_i, q)}\nonumber\\
&&~~~~ + \prod_{j=1}^{N_0}\alpha_2(q, q_j) 
\prod_{i=1}^{N_1}\frac{\alpha_1(q, p^{(1)}_i)}{\alpha_2(q, 
p^{(1)}_i)}\Lambda^{(1)}(q, \{p^{(1)}_l\}),
\end{eqnarray}
and
\begin{eqnarray}
&& \Lambda^{(m)}(q, \{p^{(m)}_l\}, \{p_k^{(m+1)}\}) = 
\prod_{k=1}^{N_{m+1}} \frac{\alpha_1(p^{(m+1)}_k, 
q)}{\alpha_2(p^{(m+1)}_k, q)}\nonumber\\
&&~~~~~~~~ + \prod_{l=1}^{N_m}\alpha_2(q, 
p^{(m)}_l)\prod_{k=1}^{N_{m+1}}\frac{\alpha_1(q, p^{(m+1)}_k)}{\alpha_2(q,
p^{(m+1)}_k)}\nonumber\\
&&~~~~~~~~ \times \Lambda^{(m+1)}(q, \{p^{(m+1)}_k\}),
\end{eqnarray}
where $m = 1, 2, \cdots, 2n-3$. We have defined $N_{2n-2}=0$, and
$\Lambda^{(2n-3)}(q, \{p^{(2n-2)}_l\}) = 1$.

We also have the following $(2n-3)$ sets of Bethe ansatz equations:
\begin{eqnarray}
&& \prod_{j=1}^{N_m}\frac{\alpha_1(p^{(m+1)}_i,  
p^{(m)}_j)}{\alpha_2(p^{(m+1)}_i, p^{(m)}_j)} = 
\prod_{l=1}^{N_{m+2}}\frac{\alpha_1(p^{(m+2)}_l, 
p^{(m+1)}_i)}{\alpha_2(p^{(m+2)}_l, p^{(m+1)}_i)}\nonumber\\
&&~~~~~~~~~~~~~~~ \times \prod_{k=1, k \neq
i}^{N_{m+1}}\frac{\alpha_2(p^{(m+1)}_k, 
p^{(m+1)}_i)}{\alpha_2(p^{(m+1)}_i, p^{(m+1)}_k)},
\end{eqnarray}
where $m = 1, 2, \cdots, 2n-3$ and the periodic boundary condition
(\ref{eigenvalue}) implies:
\begin{equation}
e^{i k_j L} = \prod_{l=1}^{N_1}\frac{\alpha_1(p^{(1)}_l,
q_j)}{\alpha_2(p^{(1)}_l, q_j)},
\end{equation}
so we get $2(n-1)$ sets of Bethe ansatz equations.

After Substituting the expressions of $\alpha_1, \alpha_2$ into above
expressions, the $2(n-1)$ sets of Bethe ansatz equations become to be
\begin{equation}
\label{first-BA}
e^{i k_j L} = \prod_{l=1}^{N_1}\frac{(\tilde{p}^{(0)}_j - 
\tilde{p}^{(1)}_l - i\frac{\gamma}{2})}{(\tilde{p}^{(0)}_j -
\tilde{p}^{(1)}_l + i \frac{\gamma}{2})},
\end{equation}
and
\begin{eqnarray}
\label{eq-BA}
&& \prod_{k=1, k \neq i}^{N_{m+1}}\frac{(\tilde{p}^{(m+1)}_i - 
\tilde{p}^{(m+1)}_k + i\gamma)}{(\tilde{p}^{(m+1)}_i - \tilde{p}^{(m+1)}_k
- i \gamma)} = \nonumber\\
&& ~~~~~~~~~~~~ \prod_{j=1}^{N_m}\frac{(\tilde{p}^{(m+1)}_i -
\tilde{p}^{(m)}_j + i\frac{\gamma}{2})}{(\tilde{p}^{(m+1)}_i -
\tilde{p}^{(m)}_j - i\frac{\gamma}{2})}\nonumber\\
&& ~~~~~~~~ \times \prod_{l=1}^{N_{m+2}}\frac{(\tilde{p}^{(m+1)}_i - 
\tilde{p}^{(m+2)}_l + i \frac{\gamma}{2})}{(\tilde{p}^{(m+1)}_i - 
\tilde{p}^{(m+2)}_l - i \frac{\gamma}{2})},\nonumber\\
& & ~~~~~~~~~~~(m=1, 2, \cdots, 2n-3).
\end{eqnarray}
Here $\tilde{p}^{(0)}_j = \sin k_j$ and we have used the shifted 
parameters $\tilde{p}^{(k)}_l = p^{(k)}_l + i k \frac{\gamma}{2}$ to bring
our equations to more symmetric forms.

In principle, we can consider the same system with twisted boundary
condition, similar as done in $SU(2)$ Hubbard model. The Bethe ansatz
equations and the energy will pick up an appropriate factor which related
to a twisted angles. On the other hand, we can also study this method
based on the fermionic base, which can be obtained by using the
Jordan-Wigner transformation from bosonic one, but the boundary condition
will be changed. Generally, the bosonic $SU(n)$ model with periodic
boundary will correspond to a fermionic one with twisted boundary and
viceversa, the fermionic system in the presence of periodic boundary
correspond to a twisted bosonic version. This is a direct generalization
of discussions give in Ref.\cite{RT}

\section{The ground state}

\indent

In the above section, we have obtained the final $2(n-1)$ sets of Bethe
ansatz equations of the $SU(n)$ Hubbard model. In this section, we will
use them to analyse the ground state of the model.

Under thermodynamic limits, after taking the lograrithm, the Bethe ansatz
equations (\ref{first-BA})-(\ref{eq-BA}) for the ground state change
into:
\begin{equation}
\label{rho^{(0)}}
2\pi\rho^{(0)}(k) = 1 - \cos k 
\int_{-B^{(1)}}^{B^{(1)}}\frac{4\gamma\rho^{(1)}(\Lambda^\prime)}{\gamma^2
+ 4(\Lambda^{(0)} - \Lambda^\prime)^2}d \Lambda^\prime,
\end{equation}
and
\begin{eqnarray}
\label{rho^{(m)}}
2\pi\rho^{(m)}(\Lambda^{(m)}) && +
\int_{-B^{(m)}}^{B^{(m)}}\frac{2\gamma\rho^{(m)}(\Lambda^\prime)} 
{\gamma^2 + (\Lambda^{(m)} - \Lambda^\prime)^2}d \Lambda^\prime =
\nonumber\\
& & \int_{-B^{(m-1)}}^{B^{(m-1)}}\frac{4\gamma
\rho^{(m-1)}(\Lambda^\prime)}{\gamma^2 + 4(\Lambda^{(m)} -  
\Lambda^\prime)^2}d \Lambda^\prime \nonumber\\
&& + \int_{-B^{(m+1)}}^{B^{(m+1)}}\frac{4\gamma 
\rho^{(m+1)}(\Lambda^\prime)}{\gamma^2 + 4(\Lambda^{(m)} -
\Lambda^\prime)^2}d\Lambda^\prime,\nonumber\\
& & ~~~~~~~(m = 0, 1, \cdots, 2n-3), 
\end{eqnarray}
where $\Lambda^{(0)} = \sin k$ and $B^{(m)}$ are determined by the
conditions
\begin{equation}
\label{B^{(m)}}
\int_{-B^{(m)}}^{B^{(m)}}\rho^{(m)}(\Lambda)d \Lambda = \frac{N_m}{L},\\
\end{equation}
and $B^{(2n-2)} = 0$. The functions $\rho^{(m)}(\Lambda)$ are the
distribution functions of real parameters $k_j$ and $\tilde{p}^{(i)}_l,
(i=1, 2, \cdots, 2n-3)$ respectively.

Eqs. (\ref{rho^{(0)}})-(\ref{rho^{(m)}}) have a unique solution which
is positive for all allowed $B^{(m)}$. $\frac{N_m}{L}, (m=0, 1, \cdots,
2n-3)$ are monotonically function of $B^{(m)}$ respectively. Thus the
ground state is characterized by $B^{(0)}=\pi, B^{(m)} = \infty, (m = 1,
2, \cdots, 2n-3)$.

After taking Fourier transforms of the above equations we can obtain the
result of the distribution functions
\begin{eqnarray}
&& \rho^{(0)}(k) = \frac{1}{2\pi} - \frac{\cos{k}}{2\pi}\nonumber\\ 
&& \times\int_{-\infty}^{\infty}\frac{\sinh((2n-3)\frac{\gamma}{2} 
|\omega|)e^{-\frac{\gamma}{2}|\omega|}}{\sinh((n-1)\gamma|\omega|)}
J_0(\omega)e^{-i\omega\sin{k}}d\omega,
\end{eqnarray}
and
\begin{eqnarray}
\label{rho-m}
&& \rho^{(m)}(\Lambda) = \nonumber\\  
&&~~\frac{1}{2\pi}\int_{-\infty}^{\infty}\frac{\sinh((2n - 2 -
m)\frac{\gamma}{2} |\omega|)}{\sinh((n-1)\gamma|\omega|)} J_0(\omega)
e^{-i\omega\Lambda}d\omega,\nonumber\\
&&~~~~~~~~~~~~~~~(m = 1, 2, \cdots, 2n-3).
\end{eqnarray}
From Eqs.(\ref{B^{(m)}})-(\ref{rho-m}) we have $N_0 = L$, this means
that all lattice site is filled by one particles, and $N_m = 
\frac{2n-2-m}{2(n-1)}N_0$, these corresponding to the $2(n-1)$ conserved
particle numbers.

The ground state energy then is given by
\begin{eqnarray}
E & = & 2 L \int_{-\infty}^{\infty}\rho(k)\cos{k}d k - \frac{\gamma}{2}L
\nonumber\\
& = & - 2 L
\int_{-\infty}^{\infty}\frac{\sinh((2n-3)\frac{\gamma}{2}|\omega|) 
e^{-\frac{\gamma}{2}|\omega|}}{\omega\sinh((n-1)\gamma|\omega|)} 
J_0(\omega)J_1(\omega)d\omega\nonumber\\
& &  - \frac{\gamma}{2}L.
\end{eqnarray}

\section{Conclusions}

\indent

The main purpose of this paper is to investigate the eigenvalues and the 
eigenstates of the Hamiltonian of the $SU(n)$ Hubbard model. We have
succeeded in finding the eigenvalues of the Hamiltonian
(\ref{Hamiltonian}) and obtained $2(n-1)$ sets of Bethe ansatz equations.
Based on the Bethe ansatz equations, we have found the explicit expression
of the energy and the distribution functions of the rapidities
corresponding to the ground state for positive $U$.

From the result of present paper, we know that the result in  
Ref.\cite{HPY} is not complete. The solution of the $SU(3)$ Hubbard model
obtained in Ref. \cite{HPY} was not the ground state solution. Next, from
present paper we also know that the number of Bethe ansatz equations
should be equal to the number of the conserved particle numbers. But in
Ref.\cite{HPY} one set of Bethe ansatz equations is missing.

For $SU(n)$ Hubbard model, another important question is to study the
excitation spectrum and the low tempareture thermodynamics of the model
for both positive and negative $U$. The Bethe ansatz equations given in
present paper will play an key role. As we know, the negative $U$ has a
distinguished properties from positive $U$ case. The solution structure of
Bethe ansatz equations corresponding to the ground state and excited
spectrum for attractive case $(U<0)$ are different from those for
repulsive case. It is worthy to be studied in future.

In present paper, we have given energy spectrum of the Hamiltonian
(\ref{Hamiltonian}). As an integrable model, it is also important to find
the eigenvalue of infinit number of conserved laws. The useful approach is
algebraic Bethe Ansatz method.

In Ref.\cite{Ma2}, the $L$-operator and $R$-matrix were given, therefore,
one can define the transfer matrix as done in usual Hubbard model ($SU(2)$
case). But how to find the eigenvalue of the transfer matrix is unknown.
It may be solved by using the method proposed in Ref.\cite{Martins2}. We
will consider this problem late.

\end{document}